\documentclass[epjST]{svjour}
\usepackage[sort&compress,numbers]{natbib}
\usepackage{graphicx,amsmath,amssymb,soul,bbm,slashed}
\usepackage{dcolumn,booktabs,multirow,array,float}
\usepackage{enumitem,lmodern,dsfont,nicefrac}
\usepackage[dvipsnames]{xcolor}
\usepackage[utf8]{inputenc}
\usepackage{mathtools}
\usepackage{bm,wasysym}
\usepackage[export]{adjustbox}
\usepackage{scrextend}

\usepackage[pdftex,
            pdftitle={Review $\Lambda(1405)$},
            pdfauthor={Maxim Mai},
            bookmarks,
            colorlinks,
            linkcolor=RubineRed,
            citecolor=RubineRed,
            menucolor=black,
            urlcolor=NavyBlue,
            plainpages=false,
            pdfpagelabels,
            hypertexnames=false]{hyperref}

\begin{document}

\title{REVIEW OF THE \texorpdfstring{${\mathbf \Lambda}$}{}(1405)}
\subtitle{A CURIOUS CASE OF A STRANGE-NESS RESONANCE}
\author{Maxim Mai
\inst{1}\fnmsep\thanks{\email{maximmai@gwu.edu}}}
\institute{The George Washington University, Washington, DC 20052, USA}
\abstract{
During its long-lasting history, the $\Lambda(1405)$ has become a benchmark for our understanding of the $\rm SU(3)$ hadron dynamics. Starting with its theoretical prediction and later experimental verification, until the most recent debates on the existence of the second broad pole, it emerged as a fruitful research area sparking many theoretical and experimental developments.\\
This review intends to provide the reader with the current status of research on the $\Lambda(1405)$-resonance, reflecting on historical, experimental and theoretical developments. A common database for experimental results and a comparison of most recent theoretical approaches will be provided in the last two parts of this manuscript.
}
\maketitle

\section{Introduction}
\label{sec:intro}

Excited states of strongly interacting particles form a non-trivial and highly populated spectrum. Many features of this hadronic spectrum can be understood studying the excitations of three (for baryons) or two (for mesons) constituent quarks~\cite{Capstick:1986bm,Loring:2001kx}. This picture is, however, incomplete, leading to, e.g., the so-called missing resonance problem or puzzling relative mass ordering between some states. Quantum Chromodynamics (QCD) emerged as the field theory of strong interaction successfully passing all tests for nearly half a century. Sophisticated methods have been developed to perform hadron spectroscopy from QCD leading to many valuable insights, e.g., on the spectrum of baryons~\cite{Durr:2008zz}. The excited spectrum, however, still contains riddles related to the existence of exotic states, gluonic degrees of freedom and interplay of those with with multi-hadron dynamics. It is exactly this type of riddles which challenges our understanding of strong interaction and may ultimately lead to a deeper insight into it. A prominent example of this is the enigmatic $\Lambda(1405)$ -- a $I(J^P)=0(1/2^-)$ baryonic resonance of strangeness $S=-1$. It is the matter of the present review to epitomize the history and current status of our understanding of the nature of this resonance. Previous reviews on related topics can be found in Refs.~\cite{Sakurai:1960ju, Dalitz:1963xk, Hyodo:2011ur, Gal:2016boi, Meissner:2020khl} and in the review section of Ref.~\cite{Tanabashi:2018oca}.

This review is organized in three major parts. First, an overview will be given in {\bf Section\,\ref{sec:case}}, including historical remarks, current values of $\Lambda(1405)$ resonance parameters as well as the impact of this research on nuclear and other areas of physics. In {\bf Section\,\ref{sec:experiments}} phenomenological constraints will be summarized. {\bf\,Section\,\ref{sec:theory}} will be devoted to an overview, description and comparison of theoretical approaches.

\section{The case of the \texorpdfstring{${\mathbf \Lambda}$}{}(1405)}
\label{sec:case}

\subsection{Early history}~\\[-0.8cm]
The history of the $\Lambda(1405)$ began not long after the initiation of the first large experimental programs on production of kaons in 1950's, see, e.g., Ref.~\cite{Sakurai:1960ju}. In the latter review on strong interaction, J.~J.~Sakurai spends long time discussing possible mechanisms and controversy related to the resonant structure observed previously by Dalitz and Tuan~\cite{Dalitz:1959dq}. In particular, two of four solutions of their $K$-matrix formulation constrained by the available experimental data at that time, exhibited resonance-like behavior below the $K^-p$ threshold. Indeed, two years later a resonant structure was confirmed in the hydrogen bubble chamber experiments~\cite{Alston:1961zzd,Bastien:1961zz} at $1405$~MeV in the mass ($\pi\Sigma$) plots of $K^-p\to\Sigma^{..}\pi^{..}\pi^{..}$ reactions. In parallel, $K^-$ interactions in emulsion showed peaking behavior in the $\pi\Sigma$ mass spectra at the same energy~\cite{Eiskemberg:1961}.

Similar experimental efforts led also to discoveries of further strangeness $S=-1$ resonances, e.g., $\Sigma(1385)$ $I(J^P)=1(3/2^+)$, $\Lambda(1520)$ $I(J^P)=0(3/2^-)$ and others, see Ref.~\cite{Dalitz:1963xk} for an overview of early experimental efforts. Thus, the $\Lambda(1405)$ gained its name and a permanent position (since 1963~\cite{Roos:1963zi}) in the tables of particle data group (PDG). Notably is also that the determination of the spin-parity quantum number was precluded for a long time by experimental limitations. This was finally overcome in the studies of photo-induced reactions by the CLAS collaboration~\cite{Moriya:2014kpv}, directly confirming the $J^P=(1/2^-)$ hypothesis.

\begin{figure}[t]
  \begin{center}
    \includegraphics[width=0.99\linewidth,trim=0.5cm 5cm 6cm 3.5cm,clip]{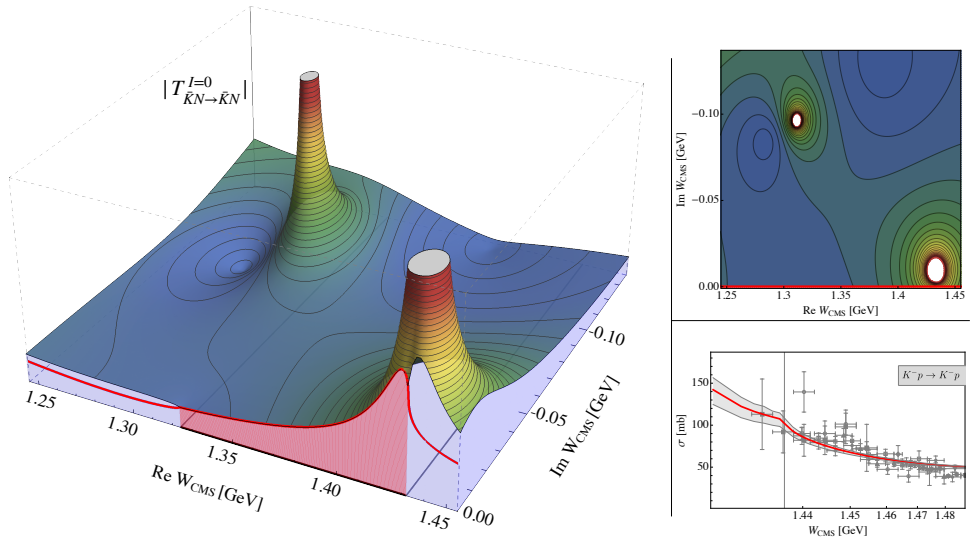}
  \end{center}
  \caption{
  \label{fig:DoublePole}
  An example of an analytic continuation of the scattering amplitude (solution 4 from Ref.~\cite{Mai:2014xna}) to the second Riemann sheet connected to the real energy-axis between the $\pi\Sigma$ and $\bar KN$ thresholds (red shaded area).  Gray lines depict the position of the $\pi\Sigma$ and $\bar KN$ thresholds in the isospin symmetric limit.
  Red line shows the behavior of the amplitude for real energies, which is constrained by the experimental data as depicted in the bottom right figure (c.f. {\bf Fig.\,\ref{fig:CSdata}}). The top right inset shows the contour plot of the same Riemann sheet, which is frequently used to visualize pole structure (c.f. {\bf Figs.\,\ref{fig:PDG-results} and \ref{fig:poles}}).
  }
\end{figure}

\subsection{Poles, mass, width}~\\[-0.8cm]
Besides discrete quantum numbers such as parity or spin, the universal parameter of a stable hadron is its mass. In the case of unstable states, the latter becomes a complex valued number quantified by the pole position of the $S$-matrix analytically extrapolated of to the complex energies. The corresponding Riemann surface consists of one physical and multiple unphysical Riemann sheets\footnote{Each Riemann sheet spans over the whole complex plane and is connected analytically to the next sheets along the cuts. In the simple case of $2\to2$ scattering of asymptotically stable particles, the cuts are located along the real axis, such that the number of Riemann sheets is $2^N$, for $N$ being the number of 2-particle thresholds.}. The poles associated with a resonance can only lie on the unphysical sheets, and are typically located on the one connected most closely to the physical one. This is because the physical information from experimental measurements or results of Lattice QCD constrain the $S$-matrix only along the real energy axis. An example of the scattering amplitude extrapolated to the second Riemann sheet is presented in {\bf Fig.\,\ref{fig:DoublePole}}.

\begin{figure}[t]
  \begin{center}
    \includegraphics[width=0.999\linewidth,trim=0 0cm 0 0cm]{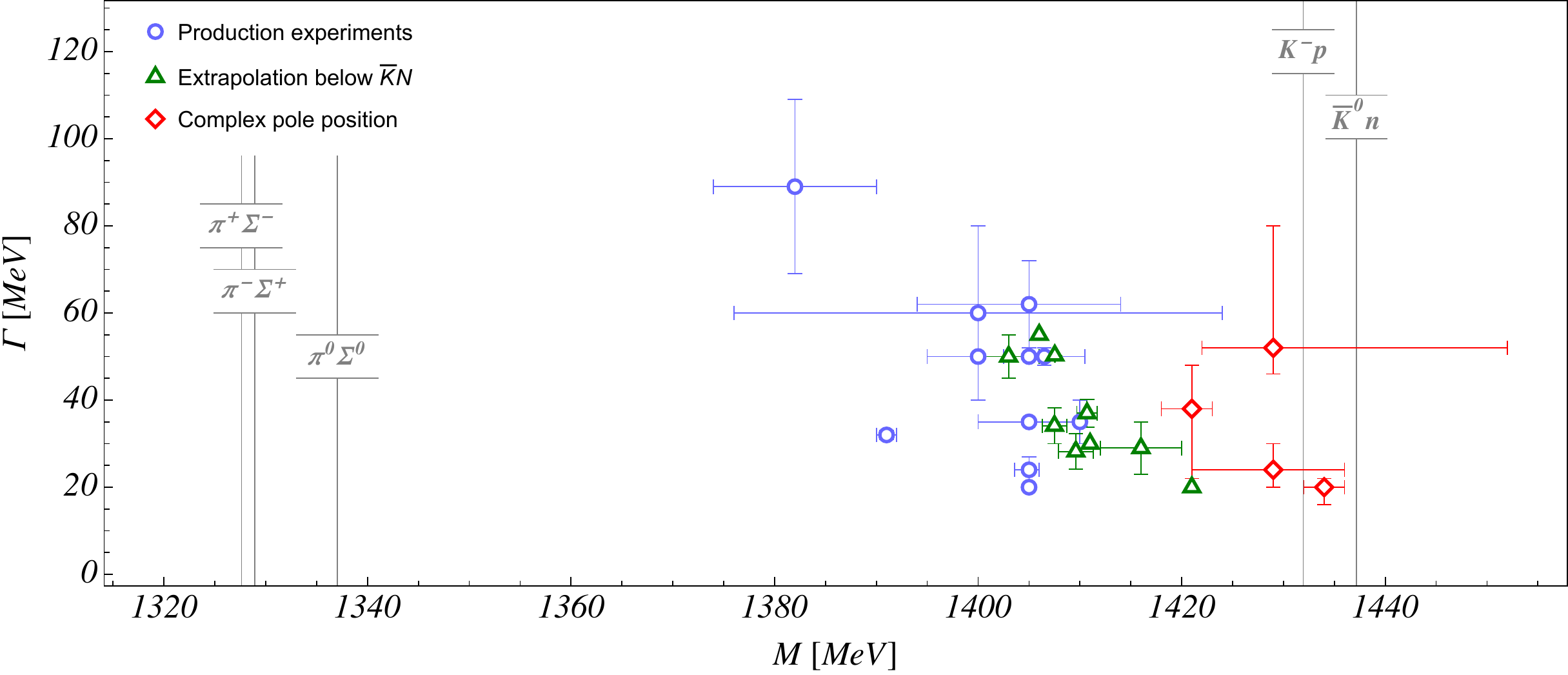}
  \end{center}
  \caption{
\label{fig:PDG-results}
Collection of results for mass vs. width of the $\Lambda(1405)$ as quoted by the PDG~\cite{Tanabashi:2018oca}, while being
separated in three groups. The \textit{``pole positions''} width and mass have been determined from the complex pole position of the narrow pole via ${z_R=(M,\Gamma/2)}$ for comparison.
}
\end{figure}

For very narrow resonances the continuation to the complex energies can be avoided by approximating the complex-valued pole position by $z_R\approx(M_R,-\Gamma/2)$ with the latter both quantities (resonance mass and width) estimated directly from the experimental line-shape. However, in the simple case of $2\to2$ scattering such data does not exists for the   $\Lambda(1405)$, which is well below the production threshold of the initial $\bar KN$ pair used to conduct scattering experiments. Thus, one is left with two choices:
(1) Use solely the scattering data above the $\bar KN$ threshold to constrain the $2\to2$ scattering amplitude. Then extrapolating below threshold and to complex energies determine the pole position of the $\Lambda(1405)$. Or (2), reduce the invariant mass of the final meson-baryon pair (e.g., $\pi\Sigma$) by introducing one or more additional particles in the final state. The lineshape with respect to the invariant mass (meson-baryon pair) allows then to extract mass and width of the $\Lambda(1405)$ approximately as discussed above. Evidently, both these choices have their advantages and are therefore contained in the most recent PDG tables~\cite{Tanabashi:2018oca} referred to as \textit{``pole positions''} and \textit{``extrapolations below $\bar KN$ threshold''} for (1), and  \textit{``production experiment''} for (2). A depiction of all quoted results is presented in {\bf Fig.\,\ref{fig:PDG-results}}. It shall be noted that due to the intricacy of the theoretical description of the $n$-body dynamics the latter case gives access only to the approximative quantities $(M_R,-\Gamma/2)$.
Accessing universal parameters of $\Lambda(1405)$ would require further development of theoretical many body tools. For most recent developments of such  methods see Refs.~\cite{Mai:2017vot, Jackura:2018xnx, MartinezTorres:2007sr} and references therein.

\subsection{Broader impact}
\label{subsec:impact}~\\[-0.8cm]
Starting with its prediction~\cite{Dalitz:1959dq} and later experimental verification, the $\Lambda(1405)$ appeared as a critical test for many theoretical tools. In chronological order three representative examples are: (1) The early debates about the attractive/repulsive nature of the (anti)kaon-nucleon interaction and the interplay of Yukawa and hypercharge current coupling, as discussed in Ref.~\cite{Sakurai:1960ju}; (2) The investigations of radiative decay of $\Lambda(1405)$ showing the importance of $q^4\bar q$ dynamics, see, e.g., Refs.~~\cite{Burkhardt:1991ms,Darewych:1985dc}; (3) The surprising observation of the double-pole structure of a dynamically generated $\Lambda(1405)$, discussed in {\bf Section\,\ref{sec:theory}}.

Besides being a critical test of our understanding of the $\rm SU(3)$ dynamics of QCD, the strong attraction of the $\bar KN$ system has a far reaching practical impact on other areas of nuclear physics. One example is the application to $\Lambda_b\to J/\psi \Lambda(1405)$ decay~\cite{Roca:2015tea} and similar processes with the finals state interaction dominated by the non-perturbative meson-baryon dynamics~\cite{Oset:2016lyh}.
Another example is the investigation of and search for the $\bar K$-nuclei -- $\bar KNN$, $\bar KKN$, $\bar KNNN$, etc., being part of large experimental programs such as, e.g., FINUDA@DA$\rm \Phi$NE~\cite{Agnello:2005qj,Agnello:2005tk,Agnello:1998vc}, DISTO@Saclay~\cite{Maggiora:2009gh,Maggiora:2001tx,Yamazaki:2010mu,Yamazaki:2008hm}. For more details on these experimental programs and relation to the theoretical predictions, see reviews~\cite{Hyodo:2011ur,Gal:2016boi}.

In a broader context, the search for $\bar K$-nuclei relates to the exploration of the in-medium properties of anti-kaons and strange nuclear matter~\cite{Mares:2020wfk,Gal:2014uua,Hrtankova:2018fjo}. One natural application of this is the study of equation of state of neutron stars (NS) in relation to the strangeness, see the comprehensive review~\cite{Gal:2016boi}. In a nutshell, the motivation for such investigations lies in the fact that compressed to multiples of nuclear matter densities, the core of neutron stars provides more than enough dense environment for appearance of kaon condensates~\cite{Kaplan:1986yq,Li:1997zb,Pal:2000pb,Lee:1996ef}, hyperons or more extreme scenarios of strange quark matter~\cite{Baym:1976yu}. However, the path between microscopic theory of hadron interactions (QCD or EFTs thereof) and properties of neutron stars is not an easy one and several challenges need to be overcome. For examples on EFT based calculations the reader is referred to Refs.~\cite{Lim:2020zvx,Hebeler:2010jx,Lim:2018bkq,Ramos:2000dq} and for multikaon systems on the lattice~\cite{Detmold:2008yn,Alexandru:2020xqf}. Currently, it is believed that -- if manifested -- the strange degrees of freedom (hyperons or kaon condensates) soften the equation of state of neutron stars~\cite{Li:1997zb,Pal:2000pb,Lonardoni:2014bwa,Hell:2014xva}. While such an effect was preferable to explain earlier astrophysical determination of $M_{\rm NS}(R_{\rm NS})$ relation (see, e.g., Ref~\cite{Li:1997zb}), it is at odds with more recent observations~\cite{Demorest:2010bx,Antoniadis:2013pzd}. Still, the fact remains that strange degrees of freedom can sizably alter the equation of state of neutron stars and must, thus, be taken seriously.

\section{State of the art: Experiment}
\label{sec:experiments}

Universal parameters of the $\Lambda(1405)$ can be accessed in a reaction independent way from analytical properties of the scattering amplitude in the complex complex energy-plane as described in the previous section. Several theoretical constraints can be made on such amplitudes, such as unitarity or low-energy behavior from ChPT. Necessarily, this does not fix the amplitude entirely, requiring for a phenomenological input constraining the parameters of such models. An overview of experimental data -- most relevant for the study of $\Lambda(1405)$ -- is the purpose of this section. In addition, since many sources of data are old and not well digitalized, the author has collected and sorted it in an open GitHub repository\footnote{\label{github}\url{https://github.com/maxim-mai/Experimental-Data/tree/master/Lambda1405}}.

\begin{figure}[t]
  \begin{flushright}
    \includegraphics[height=2.85cm,trim=0 1.2cm 0 0,clip]{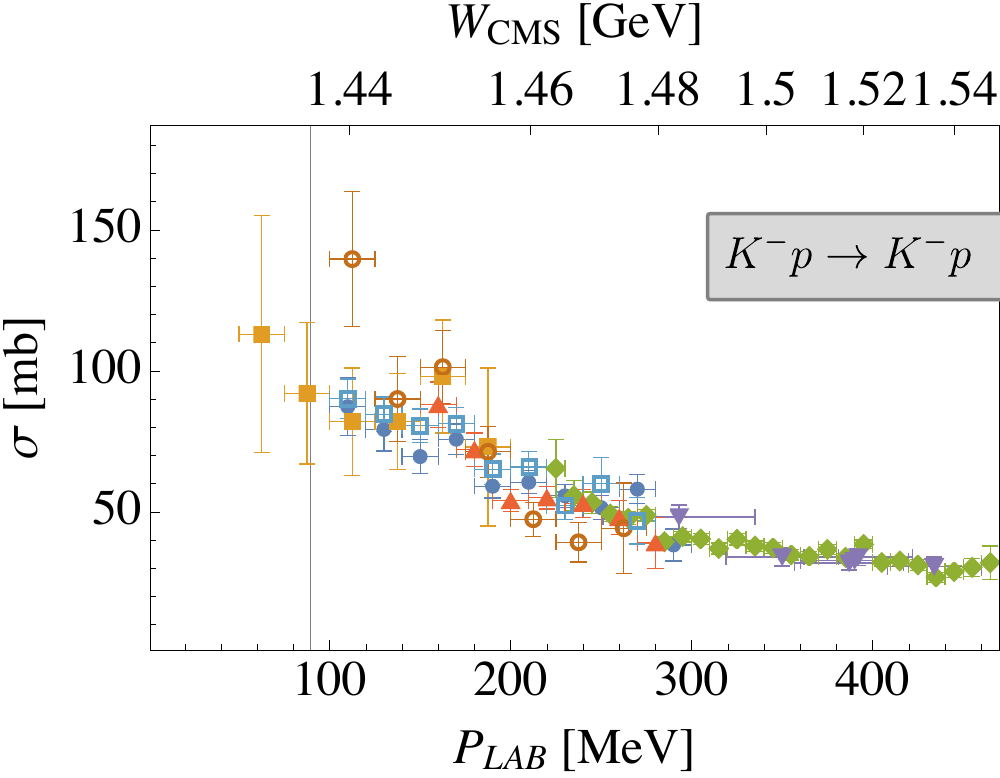}
    \includegraphics[height=2.83cm,trim=0.5cm 1.2cm 0 0,clip]{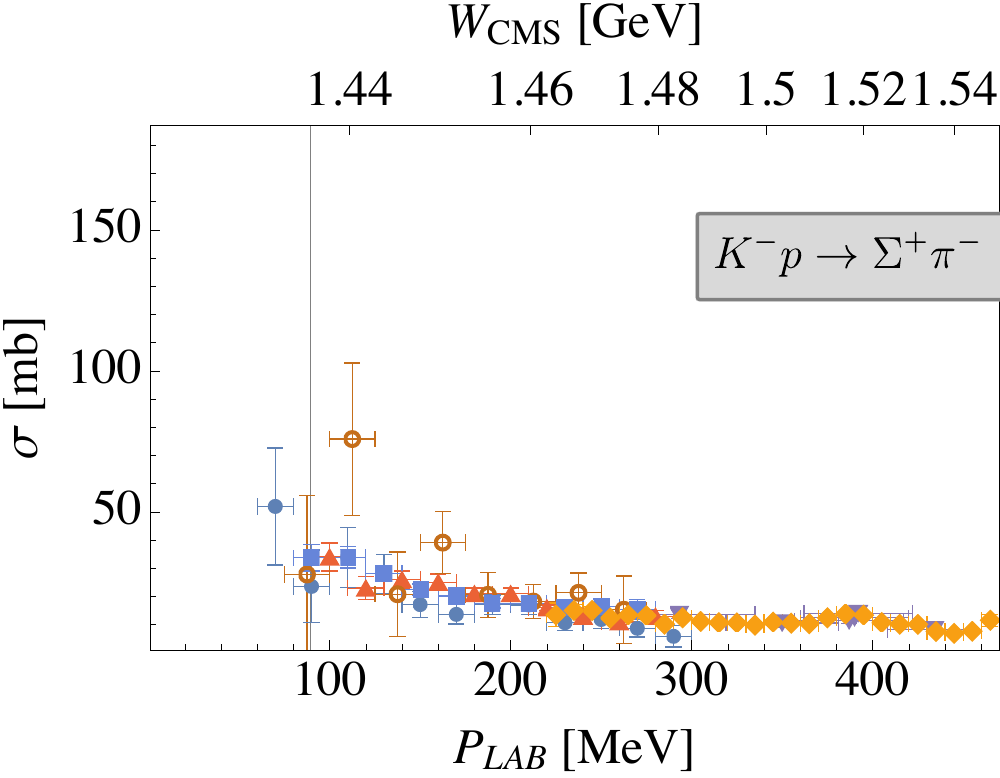}
    \includegraphics[height=2.83cm,trim=0.5cm 1.2cm 0 0,clip]{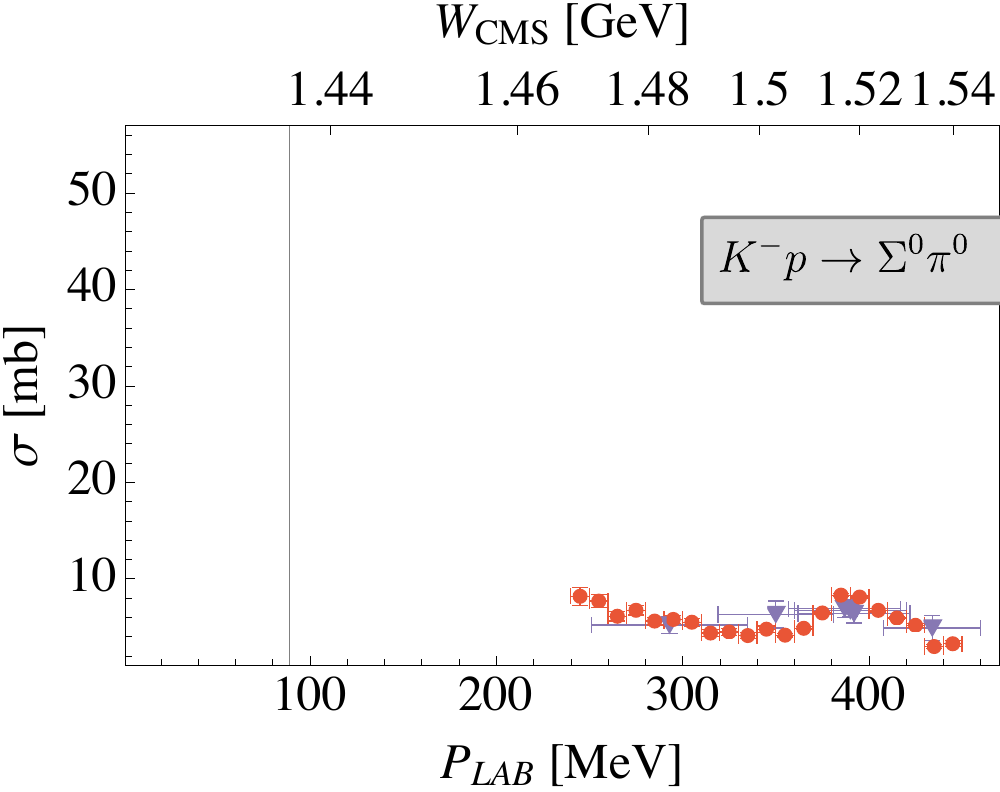}\\
    \includegraphics[height=2.83cm,trim=0 0 0 1.2cm,clip]{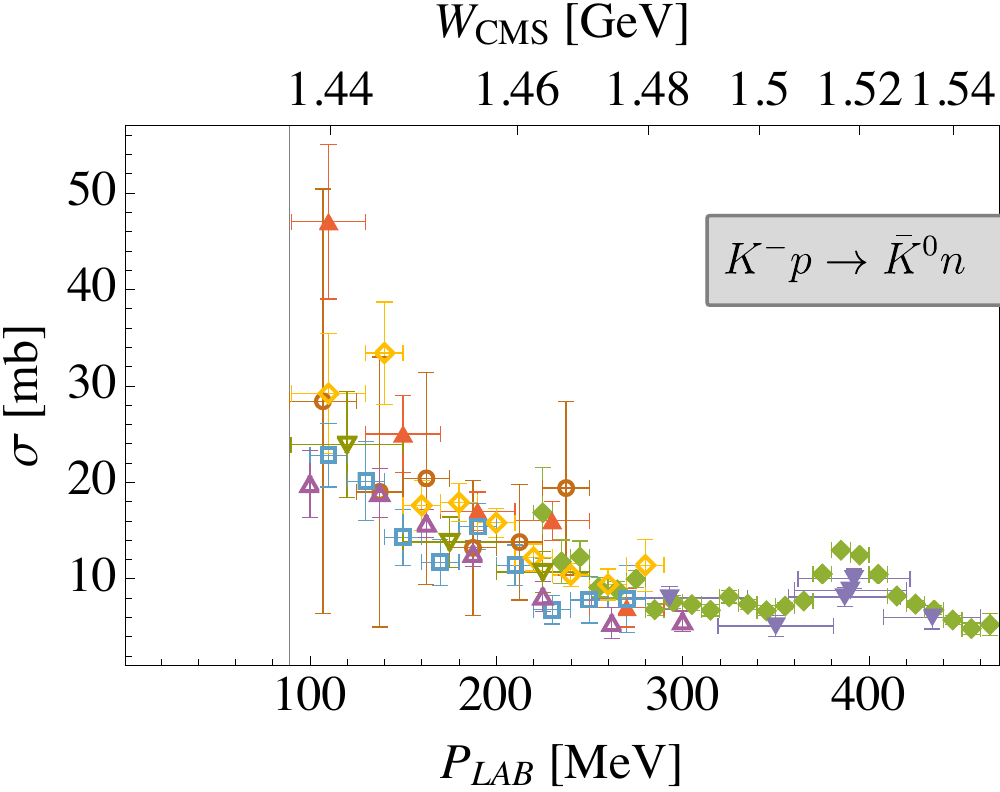}
    \includegraphics[height=2.83cm,trim=0.5cm 0 0 1.2cm,clip]{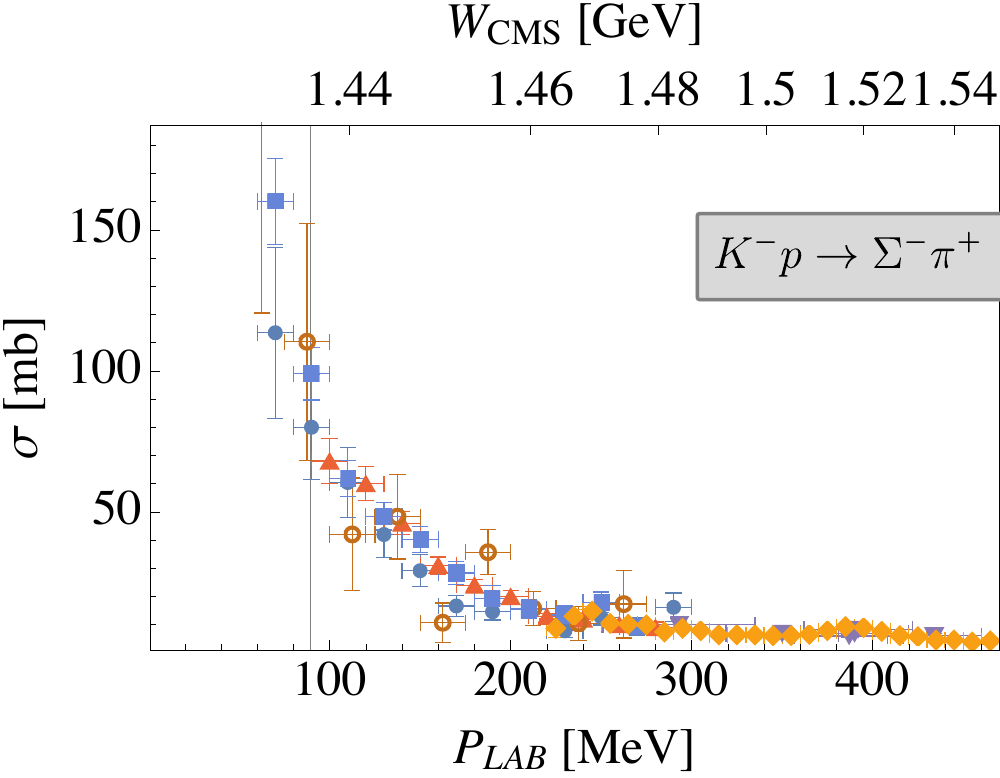}
    \includegraphics[height=2.83cm,trim=0.5cm 0 0 1.2cm,clip]{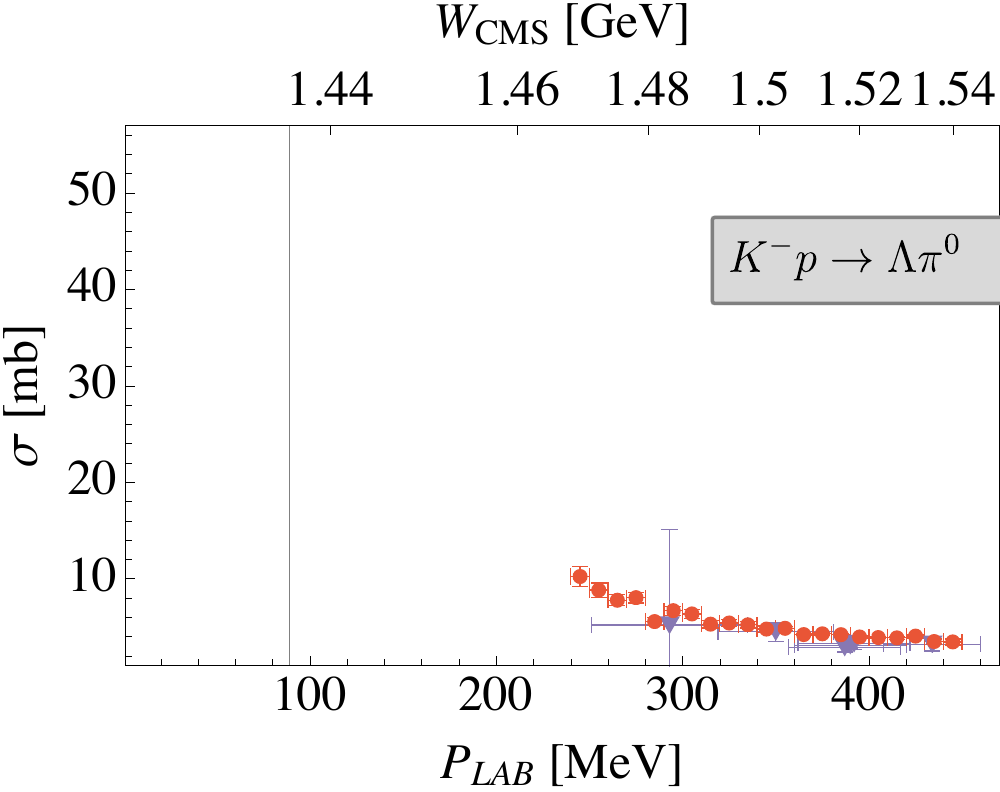}
  \end{flushright}
\begin{flushright}
\resizebox{7cm}{!}{
\begin{tabular}{rrrrrrr}
&{\color{RoyalBlue}$\bullet$}~\cite{Sakitt:1965kh}&
{\color{orange}$\blacksquare$}~\cite{Csejthey-Barth:1965izu}&
{\color{OliveGreen}$\blacklozenge$}~\cite{Mast:1975pv}&
{\color{OrangeRed}$\blacktriangle$}~\cite{Ciborowski:1982et}&
{\color{Purple}$\blacktriangledown$}~\cite{Watson:1963zz}&
{\color{Brown}$\circ$}~\cite{Humphrey:1962zz}\\
{\color{RoyalBlue}$\square$}~\cite{Kim:1967zze}&
{\color{orange}$\lozenge$}~\cite{Evans:1983hz}&
{\color{Plum}$\triangle$}~\cite{Kittel:1966zz}&
{\color{Green}$\triangledown$}~\cite{Abrams:1965zz}&
{\color{OrangeRed}$\bullet$}~\cite{Mast:1974sx}&
{\color{RoyalBlue}$\blacksquare$}~\cite{Kim:1965zzd}&
{\color{Orange}$\blacklozenge$}~\cite{Bangerter:1980px}
\end{tabular}
}
\end{flushright}
  \caption{
  \label{fig:CSdata}
  Summary of experimental data on total cross section for the reaction channels $K^-p\to\{K^-p $, $\bar K^0 n$, $\pi^{0/+/-}\Sigma^{0/-/+}\}$ from Refs.~\cite{Ciborowski:1982et,Humphrey:1962zz, Sakitt:1965kh, Watson:1963zz,Mast:1975pv,Evans:1983hz,
  Kittel:1966zz,Abrams:1965zz,Kim:1967zze,
  Csejthey-Barth:1965izu,Kim:1965zzd,Bangerter:1980px,Mast:1974sx}. Horizontal error bars represent the bin size of the corresponding measurement, while the gray vertical line shows the position of the first inelastic ($\bar K^0n$) threshold.
  }
\end{figure}

\subsection{Cross sections}~\\[-0.8cm]
The largest set of data contains total cross sections for the processes $K^-p\to\{K^-p $, $\bar K^0 n$, $\pi^{0/+/-}\Sigma^{0/-/+}\}$ measured in 1960's throughout 1980's at CERN~\cite{Abrams:1965zz,Csejthey-Barth:1965izu,Kittel:1966zz}, LBNL~\cite{Humphrey:1962zz,Sakitt:1965kh,Watson:1963zz,Mast:1975pv,Bangerter:1980px}, BNL~\cite{Kim:1965zzd,Kim:1967zze} and Rutherford Radiation Laboratory~\cite{Evans:1983hz,Ciborowski:1982et}. In those, a Kaon-beam delivered from, e.g., CERN or Bevatron was followed in a large-volume Hydrogen bubble chamber, placed in a superconducting magnet. In these impressive experimental programs, a large number of track photographs (in some cases on the order of $10^5$) were taken and evaluated for considered events using either digital techniques or ``hand analysis''. The full set of obtained data is depicted in {\bf Fig.\,\ref{fig:CSdata}}, where the energy bin sizes are represented by the horizontal error bars. For convenience of future studies the data can also be found in a digital form in the open repository\footref{github}.

%
%
%

While rather old and imprecise, these data represents the main bulk of experimental constraints on the antikaon-nucleon scattering. Additionally, some data exists on differential cross sections for elastic and charge-exchange $\bar KN$-scattering at somewhat higher $P_{\rm LAB}$~\cite{Armenteros:1970eg, Mast:1974sx, Mast:1975pv, Bangerter:1980px}. The latter data is of importance separating the hierarchy of partial waves of the scattering amplitude, see Refs.~\cite{Anisovich:2020lec,Sadasivan:2018jig}.

\subsection{Kaonic Hydrogen}~\\[-0.8cm]
A unique opportunity to gain insight into the strong $\bar KN$-dynamics is offered by the studies of the Kaonic hydrogen, with an antikaon taking the role of the electron in the hydrogen atom. Preparing such a system requires a capture of a $K^-$-meson on a proton, which demands for a very low-momentum Kaon beam. In early years of this research this was typically achieved by the emulsion techniques. Using the latter, around 3 million of kaon captures were recorded in an experiment at BNL~\cite{Tovee:1971ga}. Subsequently, around $2\permil$ decayed to $\pi^+\Sigma^-$ and $\pi^-\Sigma^+$ pairs with the ratio of branching ratios, referred to as $\gamma$. Several years later, a follow up experiment~\cite{Nowak:1978au} was conducted at the Rutherford Laboratory measuring the above as well as complimentary threshold ratios
\begin{equation}
\begin{aligned}[b]
\gamma=\frac{\Gamma_{K^-p\rightarrow
 \pi^+\Sigma^-}}{\Gamma_{K^-p\rightarrow \pi^-\Sigma^+}}\,,
\quad
R_c=\frac{\Gamma_{K^-p\rightarrow
 {\rm charged~states}}}{\Gamma_{K^-p\rightarrow {\rm all~final~states}}}\,,
\quad
R_n=\frac{\Gamma_{K^-p\rightarrow\pi^0\Lambda}}{\Gamma_{K^-p\rightarrow
\text{neutral states}}}\,.
\end{aligned}
\end{equation}
The next generation experiments~\cite{Ito:1998yi,Beer:2005qi,Bazzi:2011zj} on kaonic hydrogen were performed around 2000's. In those, similarly to the earlier experiments~\cite{Davies:1979aj,Izycki:1980uz,Bird:1983yb,Iwasaki:1997wf}, the low-energy antikaon beam was stopped in some gaseous target, measuring the X-ray emission (K-series) of the kaonic hydrogen. Compared to the electromagnetic spectrum, the measured one is shifted due to strong interaction between the $K^-$-meson and the proton. Most prominently, the energy shift and width of the 1s atomic state can be related to the complex-valued $K^-p$ (strong) scattering length $a_{K^-p}$. There are some discrepancies between the results of these experiments, which in the past led to many theoretical controverses~\cite{Weise:2010xn,Zmeskal:2008zz,Borasoy:2005fq,Oller:2005ig}. Eventually, current benchmark refers to the most recent measurement by the SIDDHARTA collaboration~\cite{Bazzi:2011zj}. Numerical values of all discussed threshold  observables are quoted below
\begin{center}
\begin{tabular}{ccccc}
$\gamma$~\cite{Nowak:1978au,Tovee:1971ga}
&$R_c$~\cite{Nowak:1978au}
&$R_n$~\cite{Nowak:1978au}
&$\Delta E$~\cite{Bazzi:2011zj}
&$\Gamma/2$~\cite{Bazzi:2011zj}\\
\toprule
$2.38\pm0.04$&
$0.664\pm0.011$&
$0.189\pm0.015$&
$283\pm42$~eV&
$271\pm55$~eV
\end{tabular}
\end{center}
but can also be found in the above mentioned GitHub repository\footref{github}.

The threshold ratios $\gamma$, $R_c$ and $R_n$ are related to the scattering amplitude by the virtue of ratios of total cross-sections in the corresponding channels. The strong energy shift and width of kaonic hydrogen is related to the complex-valued $K^-p$ scattering length ($a_{K^-p}$) via the modified Deser-type relation~\cite{Meissner:2004jr}
\begin{equation}
\begin{aligned}[b]
 \Delta E -i\Gamma/2=-2\alpha^3\mu^2_ca_{K^-p}
       \left(1-2a_{K^-p}\alpha\mu_c(\ln \alpha -1)\right)\,,
\label{eq:Deser-type}
\end{aligned}
\end{equation}
where $\alpha \simeq 1/137$ is the fine-structure constant, $\mu_c$ is the reduced mass of the $K^-p$ system. For the discussion of higher-order corrections see Ref.~\cite{Cieply:2007nv}. In the context of $\Lambda(1405)$, all four threshold kaonic hydrogen data are perhaps the most essential constraints on theoretical models for several reasons. First, these precise data lies closest to the sub-threshold energy region. Secondly, the electro-magnetic part of the meson-baryon interaction becomes important for small $P_{\rm LAB}$, but is taken care of here by using branching ratios or by the virtue of Eq.~\eqref{eq:Deser-type}, respectively. Thus, the non-trivial implementation of the Coloumb effects into the antikaon-nucleon amplitudes~\cite{Borasoy:2005ie} can be obviated, allowing one to focus solely on the strong dynamics.

\subsection{Sequential decays}
\begin{figure}[t]
  \includegraphics[height=3.2cm,trim=5cm 8cm 7.5cm 8cm,clip,valign=t]{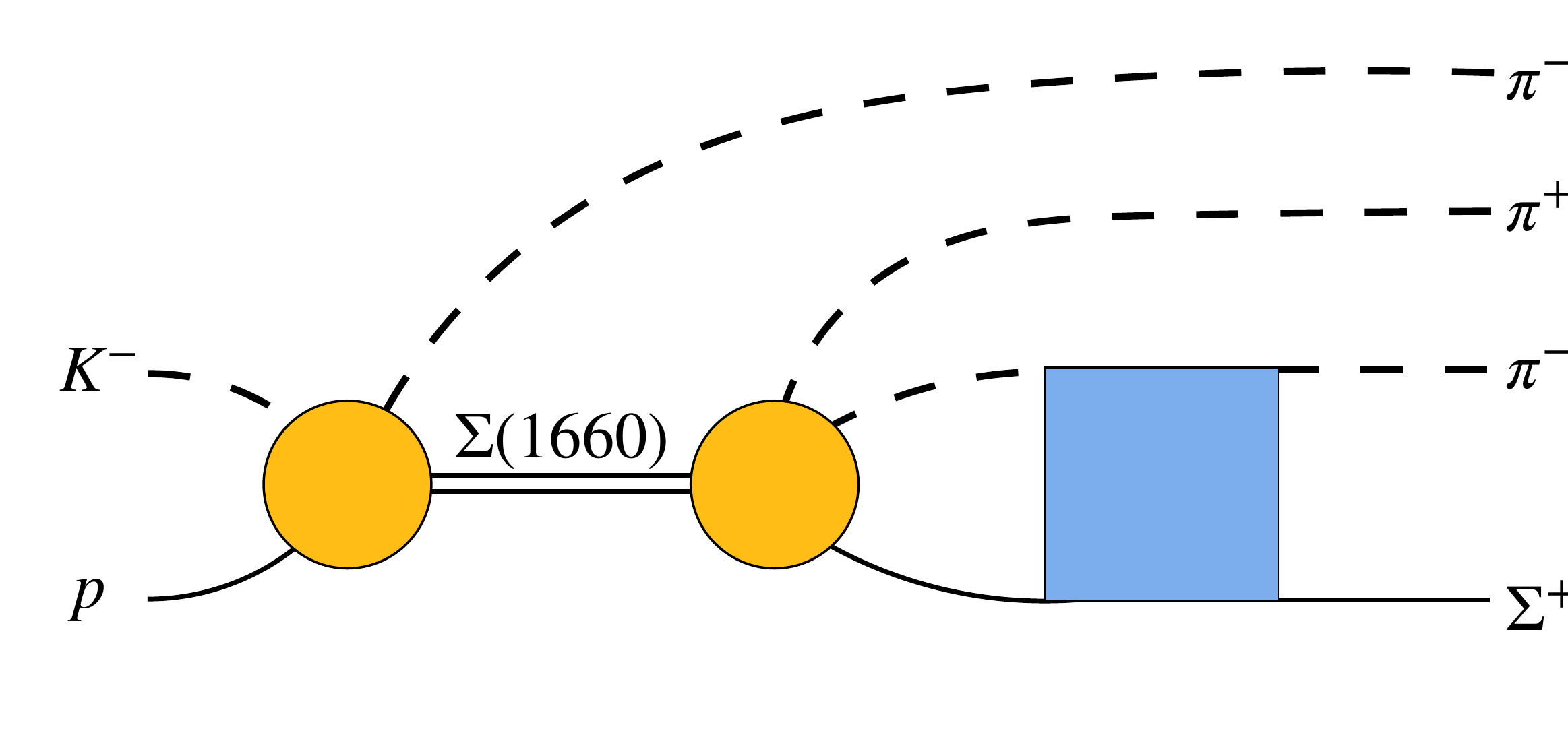}
  ~~
  \includegraphics[height=3.6cm,trim=0 0 0 0,clip,valign=t]{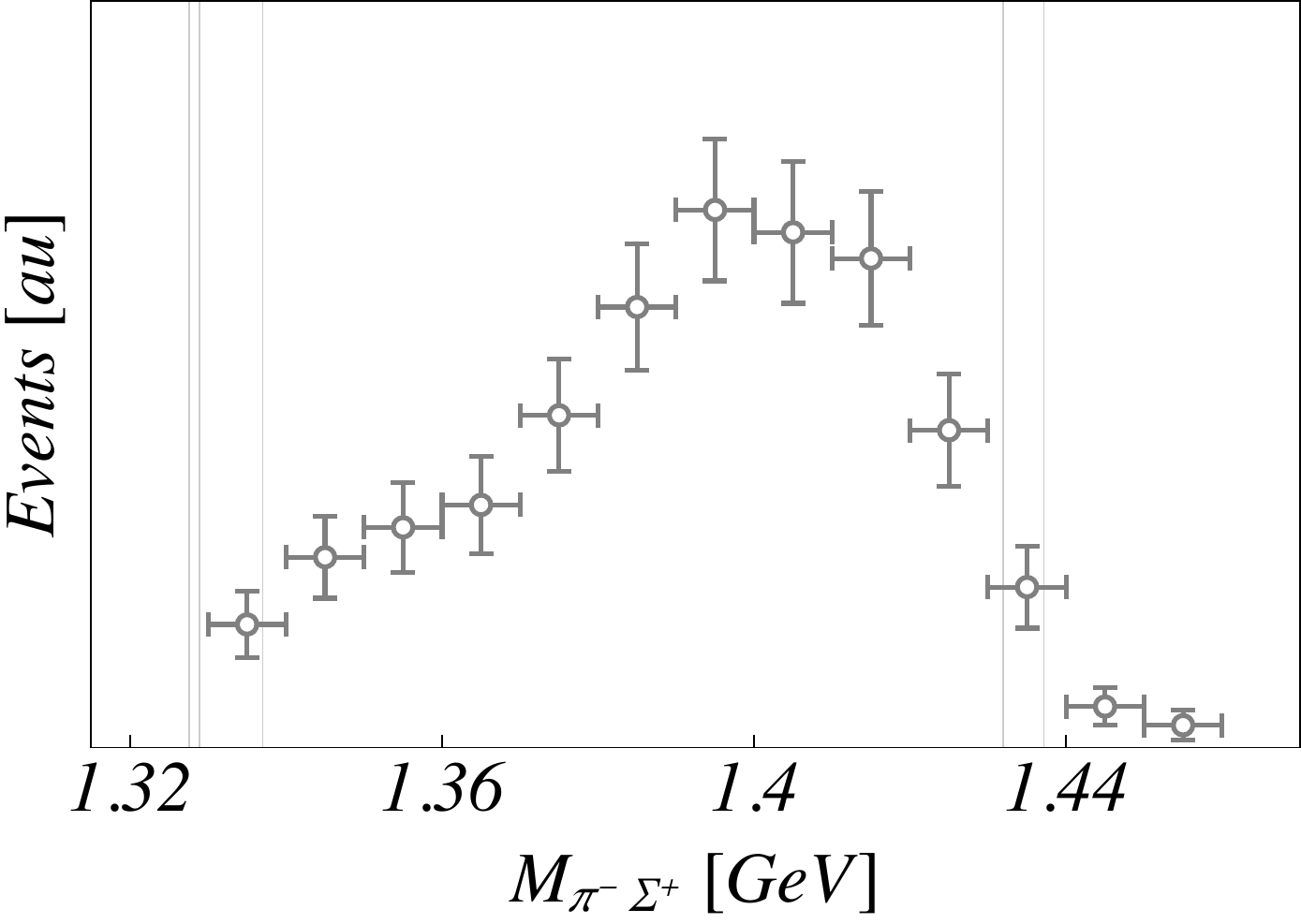}
  \caption{
  \label{fig:Hemingway}
  Production and sequential decay of $\Sigma(1660)\to\pi^-\pi^+\pi^-\Sigma^+$ asymptotic states~\cite{Hemingway:1984pz}.\newline
  {\bf Left:} Reaction mechanism with the double emission of a spectator pion (orange circles) and meson-baryon interaction of the final $\pi^-\Sigma^+$-state (blue square).\newline
  {\bf Right:} Invariant mass distribution the final $\pi^-\Sigma^+$ pair as measured in Ref.~\cite{Hemingway:1984pz}. Vertical gray lines show the positions of $\pi\Sigma$ and $\bar KN$ thresholds.
  }
\end{figure}

~\\[-0.8cm]
As discussed in the introduction, an alternative way in accessing the sub-threshold energy region of the $\bar K N$ scattering amplitude is to study multi-particle final decay states similar to the original experiments of Refs.~\cite{Alston:1961zzd, Bastien:1961zz, Eiskemberg:1961}. In such a setup, additional (to the meson-baryon pair) particles carry finite momentum away, such that one is able to probe the meson-baryon system at lower energies than the $\bar KN$ threshold. An important set of data in this context consists of the invariant-mass distribution of the $\pi\Sigma$ sub-system of the $K^-p\to\Sigma^+(1660)\pi^-\to\pi^-(\pi^+(\pi^-\Sigma^+))$ process measured in the bubble chamber experiment at CERN~\cite{Hemingway:1984pz}. Here, the meson-baryon state in the innermost parenthesis couples also to the $\Lambda(1405)$ allowing to scan for the corresponding ``bump'' in the line-shape, as depicted in {\bf Fig.\,\ref{fig:Hemingway}}. The corresponding numerical data is collected in the open GitHub repository\footref{github}.

The apparent advantages of such an experiment are out weighted by the substantially increased theoretical complexity in accessing the universal parameters -- the complex poles position and residuum of the $\Lambda(1405)$. In particular, analytically unambiguous $2\to4$ transition amplitudes are not known, such that more approximative phenomenological approaches are applied. At the very least, this yields multiple new parameters (complex- or real-valued production vertices). Given only 12 data point with somewhat low energy resolution and statistics, this leads to only very soft constraints on the theoretical $\bar KN$ models, see, e.g., Refs.~\cite{Mai:2012dt,Mai:2014xna,Anisovich:2020lec} for quantitative examples.

\subsection{Photoexcitation}~\\[-0.8cm]
The most recent experimental progress has been achieved by the CLAS collaboration measuring the $\gamma p\to K^+\Sigma \pi$ transition~\cite{Moriya:2014kpv, Moriya:2013eb} in the dedicated experiment at the Jefferson Laboratory. As mentioned before, this experiment~\cite{Moriya:2014kpv} unambiguously confirmed the spin and parity of the $\Lambda(1405)$ as $J^P = 1/2^-$, in agreement with theoretical expectations~\cite{Sakurai:1960ju,Dalitz:1963xk}. Equally important is the corresponding high precision measurement~\cite{Moriya:2013eb} of the full Dalitz plot of the final $K^+\Sigma^+\pi^-$, $K^+\Sigma^-\pi^+$ and $K^+\Sigma^0\pi^0$ systems and line-shapes of all three $\pi\Sigma$ pairs at multiple total energies.

The final state of the mentioned reaction consists of three hadrons, which makes the construction of data analysis tools cumbersome. Guided by the experience in the two-body sector, many modern methods in constructing such tools rely on unitarity as guiding principle, see e.g., Refs.~\cite{Mai:2017vot,Jackura:2018xnx,Kamano:2011ih} with some recent applications~\cite{Kamano:2008gr,Sadasivan:2020syi}. In the case of $\gamma p\to K^+\Sigma \pi$ the additional complication arises as one needs to include $S=-1$ meson-baryon sub-channel up to relatively high energies coherently to that of the $K\pi\to K^*\to K\pi$ channel. Nevertheless, a first phenomenological analysis of the Dalitz plots has been performed in a framework based on Bonn-Gatchina partial-wave analysis program~\cite{Anisovich:2020lec,Anisovich:2019exw}. Furthermore, constraints on the chiral unitary models from the $\pi\Sigma$ line-shapes~\cite{Moriya:2014kpv} have been studied previously in Refs.~\cite{Mai:2014xna, Roca:2013av}. There, a structureless ansatz was employed for the initial production vertex $\gamma p\to K^+(MB)$ with $(MB)$ denoting meson-baryon channels. As demonstrated in these works, the CLAS data indeed can reduce the model space substantially, impacting even the pole structure of
the $\Lambda(1405)$.

\subsection{Future facilities}~\\[-0.8cm]
Two types of near future experiments are expected to become the driving force for the further progress of the field. First, there is an ongoing effort for an upgrade of the so-important SIDDHARTA $\bar KH$ experiment~\cite{Bazzi:2011zj} to measure the spectrum of the $\bar Kd$-system -- the SIDDHARTA-2 experiment~\cite{Curceanu:2020vjj,Zmeskal:2019ksw}. A comparable proposal exists for an experiment at J-PARC~\cite{Hashimoto:2019qfy,Zmeskal:2015efj}. When measured, the $\bar Kd$ threshold amplitude can be related to the $\bar K N$ amplitudes directly within a non-relativistic effective field theories as derived in Ref.~\cite{Chand:1962ec,Kamalov:2000iy,Meissner:2006gx,Doring:2011xc,Mai:2014uma} or via the non-relativistic three-body Faddeev framework~\cite{Torres:1986mr,Toker:1981zh,Bahaoui:2003xb,Shevchenko:2011ce,Shevchenko:2012np,Shevchenko:2014uva}. The importance of this complementary measurement lies in the fact that only a combination of both $\bar Kd$ and $\bar KH$ measurements can resolve both Isospin contributions of the $\bar KN$ scattering amplitude at the threshold unambiguously. For example, current theoretical models agree supremely, when projected to the $K^-p$ channel, but exhibit large disagreements in the complimentary
$K^-n$ channel, see {\bf Fig.\,\ref{fig:amplitudes}}. This disagreement is concerning, but offers an opportunity to reduce the model space when the new SIDDHARTA-2 data becomes available.
Secondly, currently approved Hall D experiment~\cite{Amaryan:2020xhw} at Jefferson Laboratory intends to use secondary beam of neutral kaons performing strange hadron spectroscopy. With lowest energies of the beam of around 300 MeV, there is a possibility that the data on antikaon-nucleon cross sections can be improved, taking advantage of the isospin filtering~\cite{Feijoo:2018den}. Obviously, this is a highly desirable update of old results from bubble chamber experiments discussed before.

\section{State of the art: Theory}
\label{sec:theory}

\subsection{Theoretical context}
\label{subsec:context}~\\[-0.8cm]
Quantum Chromodynamics is the fundamental theory of the strong interaction, and must, thus, inevitably grant one an access to the properties of (excited) hadrons. However, $\Lambda(1405)$ lies at energies where the perturbative approach to QCD is of no use. Fortunately, there are tools which allow to access this energy regime in a systematic fashion, namely Chiral Perturbation Theory (ChPT) and lattice gauge theory.

ChPT~\cite{Gasser:1983yg,Weinberg:1978kz} and extensions thereof to the strangeness~\cite{Gasser:1984gg} and baryon sector~\cite{Gasser:1987rb,Bernard:1992qa,Tang:1996ca,Becher:1999he,Ellis:1997kc} have become a powerful tool and in many cases a benchmark for calculations of different observables in the threshold and subthreshold energy region~\cite{Bernard:2006gx, Bernard:2007zu, Scherer:2002tk, Meissner:1993ah, Kubis:2007iy, Bernard:1995dp,Bernard:1993fp}. However, the convergence of perturbative chiral series is impeded by the large separations of $S=-1$ meson-baryon production thresholds and the presence of the strange quark. For example, an explicit ChPT calculation of the scattering lengths in a manifestly covariant way leads to the following expansion~\cite{Mai:2009ce}
\begin{equation}
\begin{aligned}[b]
a_{\bar KN}^{I=0}=
\Big((+0.53)_{\rm LO}+(+0.97)_{\rm NLO}+(-0.40+0.22i)_{\rm NNLO}+...\Big) {\rm~fm}\,,\\
a_{\bar KN}^{I=1}=\Big((+0.20)_{\rm LO}+(+0.22)_{\rm NLO}+(-0.26+0.18i)_{\rm NNLO}+...\Big) {\rm~fm}\,.
\end{aligned}
\end{equation}
Thus, once more non-perturbative dynamics prevents one from directly accessing  the meson-baryon scattering at energies around the $\Lambda(1405)$. Of course, perturbative approach is not meaningful when accessing resonances in the first place~\cite{Savage:1994pf}. Summing up the three leading orders of the chiral expansion yields the antikaon-nucleon scattering length, which compares poorly with our best phenomenological knowledge of the latter quantity
\begin{center}
\begin{tabular}{cc}
\text{NNLO SU(3) ChPT~\cite{Mai:2009ce}}&\text{~~~~~SIDDHARTA/DTR~\cite{Bazzi:2011zj}/\cite{Meissner:2004jr}}\\
\toprule
$a_{\bar KN}^{I=0}=\big(+1.11+0.22i\big)~{\rm fm}$
&~~~~~$a_{\bar KN}^{I=0}\approx\big(-0.53+0.77i\big)~{\rm fm}$
\end{tabular}
\end{center}
Here the value quoted in the right column refers to the measurement of the energy shift and width of kaonic hydrogen in the SIDDHARTA experiment at ${\rm DA \Phi NE}$~\cite{Bazzi:2011zj}, which is related to the antikaon-nucleon scattering length by the Deser-type formula~\cite{Meissner:2004jr}.

Extracting the resonance parameters of the $\Lambda(1405)$ in the {\bf twice non-perturbative regime of QCD} is a challenging task, being faced in the past by many theoretical approaches. The most representative (pre- and post-QCD) classes\footnote{This differentiation is not unique, due to substantial systematical and historical overlap between the potential, Chiral Unitary and dynamical coupled-channel models.} of those are:
\begin{table}[h]
\centering
\begin{tabular}{p{6cm}p{6cm}}
$\bullet$ Potential models~\cite{Dalitz:1967fp,Landau:1982mi, Schnick:1987is,Siegel:1988rq, Schnick:1989vf, Fink:1989uk,Kaiser:1995eg,Cieply:2009ea, Cieply:2011nq,Shevchenko:2011ce,Hassanvand:2012dn,Revai:2017isg,Revai:2019ipq};&
$\bullet$~Cloudy bag models~\cite{Jennings:1986yg, Thomas:1981vc, Umino:1989at};\\
$\bullet$~QCD sum rules~\cite{Leinweber:1989hh,Kisslinger:2009dr};&
$\bullet$~Relativistic Quark Model~\cite{Capstick:1986bm, Darewych:1985dc};\\
$\bullet$~Bound state soliton model \cite{Schat:1994gm,Ezoe:2020piq};&
$\bullet$ Chiral Unitary Models~\cite{Oset:1997it,Oller:2000fj,Mai:2012dt,Borasoy:2005ie, Oller:2006jw,Ikeda:2011pi,Ikeda:2012au,Guo:2012vv, Borasoy:2006sr,Jido:2002zk, Sadasivan:2018jig,Cieply:2016jby,Jido:2003cb,Lutz:2001dr,Feijoo:2018den};\\
$\bullet$~Dynamical coupled-channel models~\cite{Fernandez-Ramirez:2015tfa,Anisovich:2019exw,Anisovich:2020lec,Zhang:2013sva,Haidenbauer:2010ch}&
$\bullet$~Heavy Baryon ChPT with explicit resonances~\cite{Lee:1995ku};
\end{tabular}
\end{table}

The large variety of models is an important asset in estimating the systematic uncertainty in determination of universal parameters of $\Lambda(1405)$. However, addressing above approaches in detail would require an extensive discussion of the respective historical context and is beyond the scope of the present review. Thus, further discussion is focused only on currently used approaches, i.e., Chiral Unitary, Dynamical Coupled-Channel and potential models. It is also these types of models which underlie the set of $\Lambda(1405)$-parameters quoted in the current PDG tables~\cite{Tanabashi:2018oca}.

It shall also be mentioned that a non-perturbative approach to this problem from QCD is known and already applied to countless cases of hadron spectroscopy, id est the numerical calculations of Lattice QCD. Notably, the results of such calculations are not directly comparable with experimental measurements due to various technicalities, such as finite volume effects, finite lattice spacing or unphysical pion mass. Many systematical methods have been developed to overcome these challenges, see reviews~\cite{Aoki:2019cca,Briceno:2017max}. Especially for the mesonic spectrum great progress has been achieved, see, e.g., Refs.~\cite{Briceno:2016mjc,Mai:2019pqr,Fischer:2020fvl}, already approaching the physical limit and unprecedented precision. Obviously, the case of $\Lambda(1405)$ is more complicated due to the presence of baryons, noisier energy eigenlevels, or larger required operator basis~\cite{fallica_2018}. Still, first lattice calculations already exist~\cite{Menadue:2011pd}. Also extraction of infinite volume quantities is not as simple as in the former case, but some tools have been already developed~\cite{Doring:2013glu,Molina:2015uqp,Doring:2011ip,Liu:2016wxq,Lage:2009zv}. Advancing this progress will certainly lead to the next milestone of the field either based on finite-volume spectrum of the antikaon-nucleon system (see, e.g., Ref.~\cite{Bulava:2019hpz,Paul:2018yev,Morningstar:2017spu} for a recent progress report) or more direct probing of the $\Lambda(1405)$ structure~\cite{Hall:2014uca}.

\subsection{Chiral unitary approaches}
\label{subsec:chiralunitary}~\\[-0.8cm]
Extracting the resonance parameters of the $\Lambda(1405)$ while still imposing constraints from chiral symmetry of QCD has been the main motivation behind the development of the so-called Chiral Unitary models in the late 1990's. Such models rely on ideas advocated in Ref.~\cite{Dobado:1989qm,Weinberg:1990rz,Gasser:1990bv} and use unitarity as a guiding principle for construction of the scattering amplitude with chiral amplitude as a driving term including $\rm SU(3)$ coupled-channel dynamics\footnote{Considering solely ground state mesons and baryons ten meson-baryon channels $\{K^-p$, $\bar K^0 n$, $\pi^0\Lambda$, $\pi^0\Sigma^0$, $\pi^+\Sigma^-$, $\pi^-\Sigma^+$, $\eta\Lambda$, $\eta \Sigma^0$, $K^+\Xi^-$, $K^0\Xi^0\}$ carry the correct ($S=-1$, $Q=0$) quantum numbers.}.
Recovering two-body unitarity exactly, one pays the price of giving up the chiral power counting~\cite{Weinberg:1978kz} and crossing symmetry. This makes the estimation of the systematic uncertainties intricate, but efforts have been made to tackle this problem quantitatively~\cite{Epelbaum:2014efa,Cieply:2016jby}.

The practical advantage of this type of models is their predictive power, i.e., fitting  several free parameters a description of very large energy ranges becomes feasible including an extrapolation to the complex energy-plane. In doing so, the pioneering works~\cite{Kaiser:1995eg,Oset:1997it} observed clearly a sub-threshold resonance -- the dynamically generated $\Lambda(1405)$, while the next-generation study~\cite{Oller:2000fj} revealed a surprising presence of a second, broad pole. Many studies followed these pioneering approaches modifying the form of the driving term~\cite{Mai:2012dt, Borasoy:2005ie, Oller:2006jw, Ikeda:2011pi, Ikeda:2012au, Guo:2012vv, Borasoy:2006sr}, including high energy data~\cite{Feijoo:2018den}, higher partial waves~\cite{Jido:2002zk, Sadasivan:2018jig}, studying various theoretical limits~\cite{Cieply:2016jby,Jido:2003cb,Lutz:2001dr} or extending to photo-production channels~\cite{Mai:2014xna, Roca:2013av,Geng:2007vm}. While exploring the model space of this class of models, these works have steadily recorded the presence of the second pole. Thus, the {\bf double-pole structure} as predicted by the Chiral Unitary approaches, seems to be stable with respect to variations of the models or included data. As we know now, this phenomenon is common to many hadronic systems as discussed in a recent dedicated review~\cite{Meissner:2020khl}.

\begin{figure}[t]
  \centering
  \includegraphics[width=0.5\linewidth,trim=1cm 1cm 4cm 0,clip,rotate=-90]{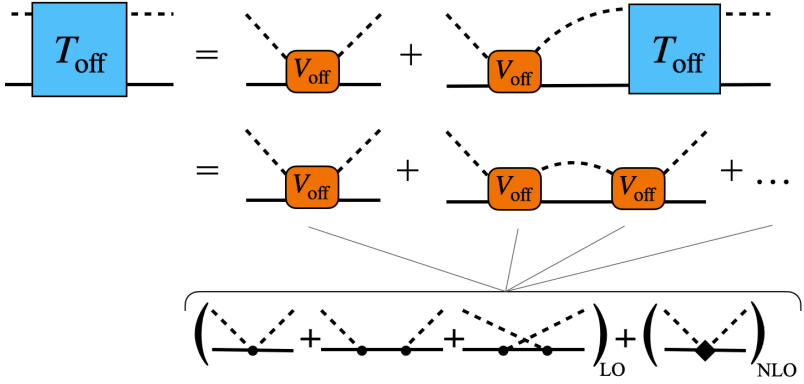}
  \caption{\label{fig:BSE}
  Diagrammatic representation of the scattering amplitude $T$ calculated in an infinite series of loop diagrams via the Bethe-Salpeter equation~\eqref{eq:BSE} with the interaction vertex $V_{\rm off}$ given in Eq.~\eqref{eq:Voff}. Dashed and full line represent meson and baryon propagators, respectively.
  }
\end{figure}

In the modern diagrammatic formulation, the Chiral Unitary approaches begin with the Chiral Lagrangian
\begin{equation}
\begin{aligned}[b]
\mathcal{L}^{\rm ChPT}_{\phi B}=\mathcal{L}^{(1)}_{\phi B}+\mathcal{L}^{(2)}_{\phi B}+\mathcal{L}^{(3)}_{\phi B}+...\,,
\end{aligned}
\end{equation}
being an infinite series of terms ordered by their chiral order (powers of meson momenta and quark masses), denoted by a subscript. The individual Lagrangians give rise to a finite set of contact terms of the type $B\to\phi B$, $\phi B\to \phi B$, etc., where $\phi$ and $B$ denote the $3\times3$ matrices containing meson and baryon fields, respectively, see, e.g., Refs.~\cite{Mai:2009ce,Frink:2006hx,Krause:1990xc}.
The crucial point is that for the discussed process of $\phi(q) B(p)\to \phi(q') B(p)$, the number of independent structures grows rapidly with the chiral order, each being accompanied by an unknown low-energy constant (LEC). Thus, for any practical calculation, the above series needs to be truncated. This differentiates between various Chiral Unitary approaches, which rely either on the leading (LO)~\cite{Oset:1997it,Oset:2016wgm,Roca:2013av} or next-to-leading (NLO)~\cite{Mai:2012dt, Borasoy:2005ie, Oller:2006jw, Ikeda:2011pi, Ikeda:2012au, Guo:2012vv, Borasoy:2006sr,Jido:2002zk, Sadasivan:2018jig} order interaction term. In the full off-shell form, the latter reads
\begin{equation}
\begin{aligned}[b]
	V_{\rm off}(\slashed{q'},\slashed{q}; P)
  =
  &\Bigg(A_{WT}(\slashed{q}+\slashed{q'})
  +A_{Bs}\slashed{q'}\frac{m-\slashed{P}}{s-m^2}\slashed{q}
  +A_{Bu}\slashed{q}\frac{m-\slashed{P}+\slashed{q'}+\slashed{q}}{u-m^2}\slashed{q'}
  \Bigg)_{\rm LO}+
  \\
	&\Bigg(A_{14}(q\cdot q')
  +A_{57}[\slashed{q},\slashed{q'}]
  +A_{M}
	+A_{811}\left(\slashed{q'}\left(q\cdot P)+\slashed{q}(q'\cdot P\right)\right)\Bigg)_{\rm NLO}\,,
\label{eq:Voff}
\end{aligned}
\end{equation}
where $P=p+q=p'+q'$, and $s=P^2$, $u=(p-q')^2$ are the usual Mandelstam variables. The matrices $A_{...}$ in the 10-dimensional channel-space depend explicitly on the meson decay constants and axial couplings $D$, $F$ in the leading, and on the LECs $\{b_0,b_D,b_F,b_1,...,b_{11}\}$ at the next-to-leading chiral order. 

The object $V_{\rm off}$ represents the chiral vertex, which can be used directly in constructing Feynman diagrams. Since the number of such diagrams to all chiral orders is infinite, a subset of an infinite cardinality is chosen which: (i) ensures unitarity exactly; (ii) includes the dominant chiral contributions defined in Eq.~\eqref{eq:Voff}. Technically, this is fulfilled by solving the Bethe-Salpeter equation
\begin{equation}
\begin{aligned}[t]
T_{\rm off}(\slashed{q'}, \slashed{q}; P)=
&V_{\rm off}(\slashed{q'}, \slashed{q};P)\\
&~~~+i\int\frac{d^4k}{(2\pi)^4}V_{\rm off}(\slashed{q'}, \slashed{k}; P)
       \frac{1}{\slashed{k}-m+i\epsilon}\frac{1}{k^2-M^2+i\epsilon}T_{\rm off}(\slashed{k}, \slashed{q}; P)\,.
\label{eq:BSE}
\end{aligned}
\end{equation}
In the past also a non-relativistic version of a unitary scattering amplitude was approached by using Lippmann-Schwinger equation, see, e.g., Ref.~\cite{Kaiser:1995eg}. In the above equation a summation in the channel space is performed over the intermediate meson and baryon states of mass $M$, and $m$, respectively. These correspond to their leading chiral order values, which are replaced in practical calculations by the physical (dressed) values, justified by the fact that there is no exact chiral power counting for $T_{\rm off}$. Regularization of the above integral equation is performed more commonly in dimensional regularization with subtraction constants used as free parameters of the model, but also momentum cutoff techniques have been applied in the past.

An analytical solution of the $T_{\rm off}(\slashed{q'}, \slashed{q}; P)$ for the case of $V_{\rm off}$ consisting only of contact interactions was found in Refs.~\cite{Bruns:2010sv,Mai:2013gwq}. In that, neglecting the $s$-channel Born diagram (last two terms in the LO-parentheses of Eq.~\eqref{eq:Voff}) was motivated by the fact that physical masses are used in the propagators of Eq.~\eqref{eq:BSE}, and that dressing the ``Lagrangian'' values of the baryon masses already includes an infinite series of iterated $s$-channel Born diagrams. Furthermore, and given the large variety of independent structures of the NLO contributions, the expectation -- backed by the findings of Refs.~\cite{Mai:2014xna,Mai:2012dt} --  is that the effects due to the $u$-channel Born term can be mimicked by the contact terms. The same studies revealed additionally that the off-shell effects due to meson-baryon intermediate channels impact the description of the data and prediction of pole-positions of $\Lambda(1405)$ only slightly. This supports the use of the on-shell condition -- applying Dirac equation and, thus, reducing the number of independent structures in Eq.~\eqref{eq:Voff}, see Ref.~\cite{Mai:2017vot}. Note that even after these modifications, the interaction term still contains the full angular structure as encoded in the NLO Chiral Lagrangian, allowing for a simultaneous description of $S-$ and $P-$ waves~\cite{Sadasivan:2018jig}.

The integral equation can be simplified further by projecting the on-shell potential to partial waves. For once, this reduces the number of relevant combinations of low-energy constants ($b_i\to d_i$) from, e.g., 14 to 7 for the $S$-wave. Furthermore, it leads to a technical advantage that the integral equation~\eqref{eq:BSE} transforms an algebraic one. Besides the possibility to incorporate the $u$-channel Born diagrams approximately, the technical simplicity of this class of Chiral Unitary led to its popularity, see, e.g., Refs.~\cite{Guo:2012vv,Borasoy:2004kk,Kaiser:1995eg,Borasoy:2005ie, Oller:2006jw, Ikeda:2011pi, Ikeda:2012au, Borasoy:2006sr}. The price to pay for this simplifications is the loss of direct connection to a series of Feynman diagrams. Instead, the formalism approaches the philosophy underlying the potential models~\cite{Dalitz:1967fp, Siegel:1988rq,Schnick:1989vf,Schnick:1987is,Fink:1989uk,Landau:1982mi,Cieply:2009ea}, but using chiral symmetry to constrain the form of the potential.

\begin{figure}[t]
  \begin{center}
  \includegraphics[width=\linewidth,trim=2cm 7cm 11cm 6cm,clip]{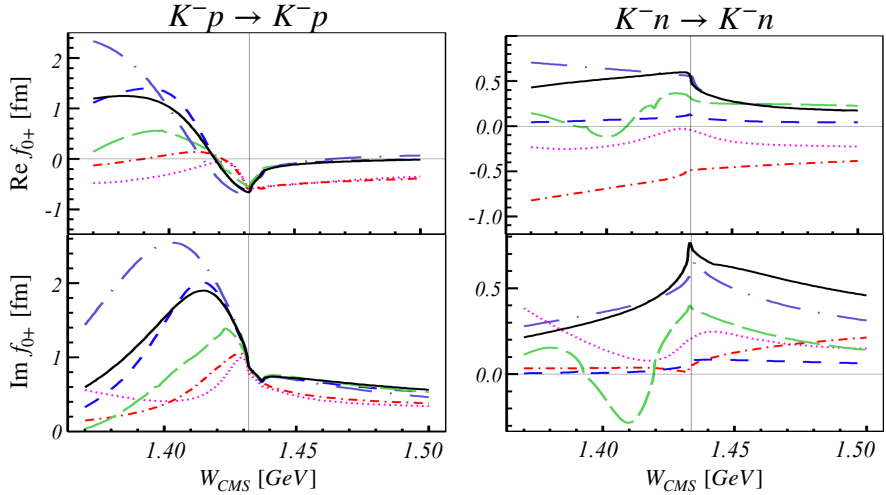}
  \end{center}
  \caption{\label{fig:amplitudes}
  Comparison of Chiral Unitary~\cite{Guo:2012vv,Mai:2017vot,Hyodo:2011ur} (colored lines) and potential model~\cite{Cieply:2011nq} (black line) predictions for the $S$-wave amplitude. Left (right) figures shows the elastic $K^-p$ ($K^-n$) channels with gray vertical line denoting the position of the $\bar KN$ threshold in the isospin limit. Original figure and further details can be found in Ref.~\cite{Cieply:2016jby}.
  }
\end{figure}

\begin{figure}[t]
  \begin{minipage}{0.6\linewidth}
  \includegraphics[width=\linewidth,trim=3.6cm 3cm 8.7cm 3cm,clip]{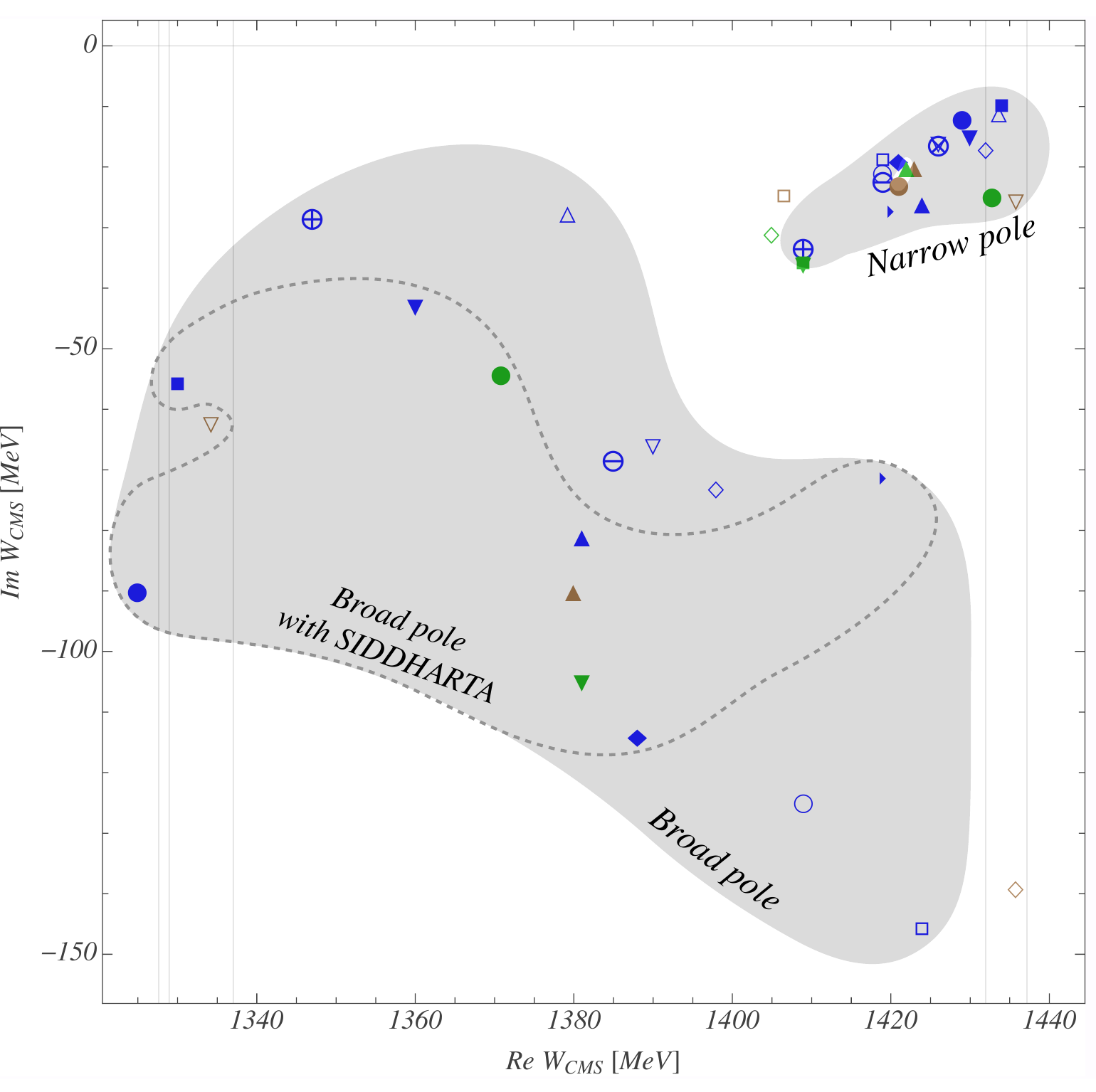}
  \end{minipage}
  \begin{minipage}{0.2\linewidth}
{\scriptsize
\begin{tabular}{ll|ll}
\multicolumn{4}{c}{Chiral Unitary Approaches}\\
\hline
\multicolumn{4}{c}{}\\
{\color{blue}$\blacksquare$}/{\color{blue}\large$\bullet$}&Ref.~\cite{Mai:2014uma}&
  \color{blue}{$\bigcirc$}&Ref.~\cite{Borasoy:2005ie}\\
\color{blue}{$\blacktriangle$}&Ref.~\cite{Ikeda:2011pi}&
  \color{blue}{$\triangle$}&Ref.~\cite{Oller:2000fj}\\
\color{blue}{$\blacktriangledown$}&Ref.~\cite{Sadasivan:2018jig}&
  \color{blue}{$\triangledown$}&Ref.~\cite{Jido:2003cb}\\
\color{blue}{$\blacklozenge$}&Ref.~\cite{Guo:2012vv}&
  {\color{blue}$\square$}/\color{blue}{$\lozenge$}&Ref.~\cite{Morimatsu:2019wvk}\\
\color{blue}{$\blacktriangleright$}&Ref.~\cite{Feijoo:2018den}&
  \color{blue}{$\otimes$}&Ref.~\cite{Oset:2001cn}\\
&&
  {\color{blue}$\ominus$}/{\color{blue}$\oplus$}&Ref.~\cite{Roca:2013av}\\
\multicolumn{4}{c}{}\\
\multicolumn{4}{c}{}\\
\multicolumn{4}{c}{Dynamical coupled-channel models}\\
\hline
\multicolumn{4}{c}{}\\
{\color{brown}\large$\bullet$}/{\color{brown}$\blacktriangle$}&Ref.~\cite{Anisovich:2020lec}&
  \color{brown}{$\lozenge$}&Ref.~\cite{Fernandez-Ramirez:2015tfa}\\
{\color{brown}$\square$}&Ref.~\cite{Zhang:2013sva}&
  \color{brown}{$\triangledown$}&Ref.~\cite{Haidenbauer:2010ch}\\
\multicolumn{4}{c}{}\\
\multicolumn{4}{c}{}\\
\multicolumn{4}{c}{Potential models}\\
\hline
\multicolumn{4}{c}{}\\
{\color{ForestGreen}\large$\bullet$}&Ref.~\cite{Cieply:2011nq}&
\color{ForestGreen}{$\blacktriangle$}&Ref.~\cite{Revai:2019ipq}\\
{\color{ForestGreen}$\lozenge$}&Ref.~\cite{Hassanvand:2012dn}&
{\color{ForestGreen}$\blacksquare$}/{\color{ForestGreen}$\blacktriangledown$}&Ref.~\cite{Shevchenko:2011ce}\\
\end{tabular}
}
  \end{minipage}
  \caption{\label{fig:poles}
  Comparison of pole predictions for poles of $\Lambda(1405)$ from most recent approaches, id est year $\ge2000$. Full symbols show the results of models, which incorporate the $\bar KH$ data from the SIDDHARTA experiment~\cite{Bazzi:2011zj}. The gray and dashed areas are drown to guide the eye in differentiating first (narrow) and second (broad) pole of double-pole solutions to the $\Lambda(1405)$. Vertical lines denote the position of $\pi\Sigma$ and $\bar KN$ thresholds.
  }
\end{figure}

In summary, all Chiral Unitary approaches rely on the Chiral Lagrangian, implementing driving term of interaction into a unitary formulation of the scattering amplitude. Methods to do so variate in:
(i) Using relativistic Bethe-Salpeter or non-relativistic Lipmann-Schwinger formulation of the scattering amplitude;
(ii) Projecting the driving term on-shell or onto specific partial waves;
(iii) Truncating the driving term to the leading or next-to leading chiral order;
(iv) Including all ten channels of ground state mesons and baryons or only the lightest six into the coupled-channel problem;
(v) Regularization of the integral equation.
As discussed before the large variety of different versions of the formalism is crucial to asses the systematic uncertainty in making one or another set of assumptions. Obviously, the largest differences occur in the regions unconstrained by the experimental data, as demonstrated in Ref~\cite{Cieply:2016jby}, comparing most recent Chiral Unitary approaches~\cite{Guo:2012vv,Mai:2017vot,Hyodo:2011ur}. For example, the prediction of the scattering amplitudes overlaps only in the energy region constrained by the experimental data as demonstrated explicitly in {\bf Fig.\,\ref{fig:amplitudes}}. The ambiguities of the $I=1$ scattering amplitudes, apparent from the right panel of the latter figure emphasize again the need for the kaonic deuterium experiments, such as SIDDHARTA-2. Besides this, the double pole structure is stable across the variations among the Chiral Unitary approaches as visualized in {\bf Fig.\,\ref{fig:poles}}.

\subsection{Potential and dynamical models}
\label{subsec:potentialmodels}~\\[-0.8cm]
Historically, the very broad class of potential models precedes the development of the Chiral Unitary approaches. It relies on some form of a potential
when solving the Lippmann-Schwinger equation
\begin{equation}
\begin{aligned}[t]
T_{\rm LS}(q',q;E)=
&V(q',q)+i\int dkk^2V(q',k)G_E(k)T_{\rm LS}(k,q;E)\,,
\label{eq:LSE}
\end{aligned}
\end{equation}
where $E$ is the total energy of the system, and $G_E$ is the  Green's function. The choice of the potential ($V$) is often motivated by its regulating properties and flexibility when implemented into the integral equation, e.g., separable potentials with Yamaguchi form factors~\cite{Fink:1989uk,Schnick:1989vf,Schnick:1987is,Cieply:2009ea,Cieply:2011nq}. In some cases phenomenological or theoretical constraints are implemented, e.g., vector-exchange potentials~\cite{Dalitz:1967fp,Siegel:1988rq} or matching to chiral potentials~\cite{Cieply:2009ea,Cieply:2011nq}.

Obviously, the potential models overlap strongly with the Chiral Unitary approaches. The main differences being besides the diagrammatic correspondence of the latter (c.f. Eq.~\eqref{eq:BSE} and {\bf Fig.\,\ref{fig:BSE}}) the relativistic effects. Note, that inclusion of relativistic kinematics is only part of such effects, and can be included into potential models~\cite{Cieply:2009ea,Fink:1989uk}. These approximations simplify the form of integral equations and still can be justified in the low energy region, c.f., full black line in {\bf Fig.\,\ref{fig:amplitudes}}.

There are two main advantages to the use of the potential models in the above sense. First, a very far-reaching impact of the research on antikaon-nucleon scattering is the study of properties of strange nuclear matter, see {\bf Section\,\ref{subsec:impact}}. Such calculations~\cite{Gal:2014uua, Cieply:2011fy, Cieply:2011yz, Cieply:2011yz, Cieply:2015pwa} profit enormously from the use of separable potential forms, which cover off-shell regions needed for the in-medium applications. Second, as a  flexible tool it allows to illuminate the impact of implemented approximations. A recent example~\cite{Revai:2017isg}, studies the form of the derived potential in relation to the off-shell effects and chiral symmetry at the leading order, discussing also the double-pole structure of the $\Lambda
(1405)$. This led to a follow-up study~\cite{Bruns:2019bwg}, which revealed stark conflicts between the latter model and chiral symmetry constraints. In offering an improvement it provides an approach, which leads again to the two-pole solution. Similar observation was made in Ref.~\cite{Morimatsu:2019wvk}, but also within the diagrammatic version of the Chiral Unitary approach~\cite{Mai:2009ce} the off-shell effects have been studied with no effect on the double-pole scenario as well.

Re-examining previously accepted approximations is crucial for further development of the field. Besides the latter studies, the dynamical coupled-channel models are important to independently test our understanding of the antikaon-nucleon scattering. Such models rely on the basic principles of scattering theory in constructing very general parametrization of scattering and (photo-) production amplitudes~\cite{Anisovich:2011ye, Ronchen:2014cna,Kamano:2013iva}. Typically, they have a large number of free parameters fixed in fits to experimental data. Thus, the descriptive power of such models is limited to the kinematic regions covered by the experimental data, and extrapolations to further energy regions (also complex-valued energies) can only be dealt with as consistency checks. An example of such a check using techniques from machine learning can be found in a recent study~\cite{Landay:2018wgf} of the pole-content of the $K\Xi$ channel. In view of the $S=-1$ channel the detectability of the double-pole structure from the experimental data was studied within the generalized optical potential in Ref.~\cite{Myint:2018ypc}. Another interesting study was conducted recently in a series of papers~\cite{Anisovich:2020lec, Matveev:2019igl, Sarantsev:2019xxm,Anisovich:2019exw}. First, the data on $K^-p$ elastic and inelastic scattering was fitted using BnGa~\cite{Anisovich:2011ye} model parameterizing the resonant and non-resonant contributions to rather high energies. A series of updates on resonance parameters of hyperons was extracted from such fits, and in a separate work a detailed study was conducted with respect to the $\Lambda(1405)$. This work included the threshold~\cite{Nowak:1978au,Tovee:1971ga,Bazzi:2011zj} and recent CLAS photo-production data~\cite{Moriya:2013eb}. Two solutions have been found in this study: a single- and a double-pole one, where the broad pole was fixed to $z_R=(1380,-90i)~\rm MeV$. Thus, at least within this model the broad pole does not influence the description of the data strongly enough and could neither be excluded nor precisely determined.

The results of all most recent determinations (year $\ge 2000$) of the pole content of $\Lambda(1405)$ are collected in {\bf Fig.\,\ref{fig:poles}}. It demonstrates the relatively small systematic uncertainty on the position of the first (narrow) pole. More importantly, the vast majority of approaches throughout the model classes supports clearly the existence of the second (broad) pole. The position of the latter is much less restricted, but becomes much less volatile when including the most recent kaonic hydrogen~\cite{Bazzi:2011zj} and CLAS photoproduction data~\cite{Moriya:2014kpv}.

\section{Summary}
\label{sec:summary}~\\[-0.8cm]

The enigma of the $\Lambda(1405)$ starting from its prediction and experimental
verification to the surprising appearance of the double-pole structure in the complex energy-plane, has become a very fruitful testing ground for approaches to the intermediate (twice non-perturbative) energy region of QCD.

At the current stage, the implementation of constraints due to chiral symmetry of QCD into unitary form of scattering amplitude seems to demand the existence of the second pole at lower energies and deeper in the complex energy-plane. This approach relies on several well-controlled and -studied assumptions. The obtained double-pole hypothesis has been confirmed by a large number of non-redundant studies exploring various theoretical limits. Inclusion of modern photoproduction and kaonic hydrogen data led to tighter constraints on the scattering amplitudes and positions of both poles.

Recent, more data driven approaches are crucial for the critical debate and re-examination of previous assumptions. Without resolving the microscopic dynamics of the hadronic states such approaches aim simply to ask which of those are demanded by the data. However, even with the high-precision photo-production data by the CLAS collaboration the two-pole scenario seems to agree with the data.

There are two major avenues, which will foster future development of the field. First, new experimental facilities may provide a new complementary data on antikaon-nucleon channel. Most importantly the kaonic deuterium experiments at J-PARC and Frescati will allow to reduce the parameter space of currently available models, which actually disagree strongly in the description of the isovector channel. Secondly, since double-pole hypothesis seems to be tied strongly to the incorporation of QCD symmetries it is crucial to foster the next-generation of Lattice QCD calculations. Either in obtaining the finite-volume spectrum in the meson-baryon channel or more direct probing of the $\Lambda(1405)$-structure.

\begin{acknowledgement}
{\bf Acknowledgments}

The author would like to thank Jose Antonio Oller for invitation to write this manuscript, and to Ulf-G.~Mei\ss ner, M.~D\"oring, A.~Cieplý, P.~Bruns for their comments and careful reading of the manuscript. The author thanks R.~Brett for help with document preparation, and is grateful to E.~Sismanidou for support and motivation.
\end{acknowledgement}

\footnotesize
\bibliography{Lambda-review-MM.bib}

\end{document}